\documentclass[aps,prd,reprint,superscriptaddress,onecolumn]{revtex4-1}
\usepackage{amsmath,amssymb}
\usepackage[colorlinks=true,allcolors=blue]{hyperref}
\usepackage{dcolumn}
\usepackage{units}
\usepackage{graphicx}
\usepackage{float}
\usepackage{enumitem}
\usepackage{diagbox}
\usepackage{overpic}
\usepackage{wasysym}
\usepackage{comment}
\usepackage{color}
\usepackage{longtable}
\usepackage{lineno}

\allowdisplaybreaks[1]

\newcommand{\be}{\begin{equation}}
\newcommand{\ee}{\end{equation}}
\newcommand{\tref}[1] {Table~\ref{#1}}
\newcommand{\eref}[1] {Eq.~(\ref{#1})}

\newcommand{\sref}[1] {Sec.~\ref{#1}}
\newcommand{\fref}[1] {Fig.~\ref{#1}}
\newcommand{\aref}[1] {Appendix~\ref{#1}}

\newcommand{\der}{\mathrm{d}}

\newcommand{\ve}[1]{\mathbf{#1}}

\begin{document}

\title{A quick algorithm to compute an approximated\\ power spectral density from an arbitrary Allan deviation} 
\author{Fabrizio De Marchi}
\email[Corresponding author:]{fabrizio.demarchi@uniroma1.it}
\affiliation{Department of Mechanical and Aerospace Engineering, Sapienza University of Rome,   Via Eudossiana, 18, 00184 Rome, Italy}
\author{Michael K. Plumaris}
\affiliation{Department of Mechanical and Aerospace Engineering, Sapienza University of Rome,   Via Eudossiana, 18, 00184 Rome, Italy}
\author{Eric A. Burt}
\affiliation{Jet Propulsion Laboratory, California Institute of Technology, Pasadena, CA, USA  }

\author{Luciano Iess}
\affiliation{Department of Mechanical and Aerospace Engineering, Sapienza University of Rome,   Via Eudossiana, 18, 00184 Rome, Italy}

\begin{abstract}

Complex architectures for wireless communications, digital electronics and space-based navigation interlink several oscillator-based devices such as clocks, transponders and synthesizers. Estimators characterizing their stability are critical for addressing the impact of random fluctuations (noise) on the overall system performance. Manufacturers typically specify this as an Allan/Hadamard Variance (AVAR/HVAR) profile in the -integration - \emph{time} domain, yet, stochastic processes governing the noise take place in the -Fourier - \emph{frequency} domain in the shape of a Power Spectral Density (PSD) function. Both are second-moment measures of the time series, however, it is only possible to translate unambiguously \emph{from} the PSD \emph{to} the AVAR/HVAR, not vice versa, except in the case of a single noise type, which is severely limiting in real-life applications.

This note elaborates an analytical method to generate an approximated PSD expressed as a set of power-laws defined in specific intervals in the frequency domain, starting from an AVAR/HVAR expressed a set of power-laws in the time domain. The proposed algorithm is straightforward to implement, applicable to all noise types (and combinations thereof) and can be self-validated by reconstructing the corresponding AVAR/HVAR by direct calculus. We also report on its limitations of and analytical expressions of the continuous version of this algorithm. Coupling with well-established algorithms relying on the PSD for power-law noise generation, the ensuing method encompasses the capability for generating multi-colored noise in end-to-end simulations, as demonstrated hereby for NASA's Deep Space Atomic Clock.

\end{abstract}


\maketitle


\section{Introduction}\label{sect_intro}

Stochastic processes governing the noise of oscillator-based devices are of great concern to engineers and scientists relying on systems distributing high-quality timing and frequency information. An illustrative example lies in the monitoring of frequency standards onboard GNSS, the main providers of navigation and timing signals. Aside from the clock's inherent instability, periodic effects such as temperature and radiation pressure induce fluctuations on the clock signal. Moreover, artificial fluctuations can arise as a consequence of the the radiometric/optical link used in the estimation process \cite{sesia2011application}, especially when using the same data type for orbit determination and time transfer \cite{dirkx2016simultaneous}. These effects result in a convoluted noise spectrum which is difficult to model in numerical simulations.

Considering the breadth and diversity of manufacturers and users of frequency and timing signals, an IEEE special issue \cite{iee_standards} standardizes the techniques and physical quantities used to measure and characterize instabilities in instruments, across time and frequency domains. The Allan Variance (AVAR), introduced in 1966 \cite{allan1966statistics}, is the most widespread statistical measure for clock and oscillator instabilities in the \emph{time} domain. It intuitively displays the uncertainty one may expect after an integration time $\tau$ following the last synchronization, and helps to isolate the effect of spurs in the signal \cite{rubiola2008phase}. Furthermore, the AVAR and its derivatives (see \cite{riley2008} for a concise yet complete description) are straightforwardly measured for long averaging times using a time interval counter. Algorithms have been proposed \cite{zucca2005clock, Galleani_2008} that can generate clock noise directly from the ADEV, however, these do not guarantee a high degree of flexibility when dealing with a multitude of noise types. In fact, these variances cannot completely encompass the covariance properties of the noise: the AVAR, for instance, is impervious to the even-symmetry components of time signals (with respect to $t = \tau$) and the Modified VAR is insensitive to $t = 3/2 \tau$ \cite{rubiola2008phase}, which is why some noise types have the same, indistinguishable signature. 

The Power Spectral Density (PSD) must be identified for a comprehensive characterization of the dominant noise type(s), setting confidence intervals of the device under test, or making bias corrections for specific variance estimators \cite{riley2008}. This is conveniently assumed to follow (a combination) of power-laws
\be
S_y (f) = h f^ \alpha
\label{Sy}
\ee
where $f$ is the Fourier (or sideband) frequency in Hertz, $h$ the intensity coefficient and $\alpha$ is the power law exponent, which distinguishes the noise processes for integer values. The main advantage of the PSD is its deterministic nature -in the limit of an infinite sample sequence- yet its measurement is limited by the bandwidth of the signal and/or the measurement system, which is why a cutoff frequency $f_H$ is defined as the minimum of the two \cite{iee_standards}.

In light of these arguments, one may prefer to characterize clock and oscillator behaviour in both time and frequency domain, hence the need to translate between the two. Although the conversion from PSD to ADEV is uniquely defined (assuming a $f_H$) the opposite is not true. Essentially, this is an ill-posed problem \cite{greenhall1997}. A direct conversion may only take place in the case of a single slope of the time variance in log-log plane, corresponding to a single noise exponent $\alpha$, which is is a constricting simplification for systems affected by a multitude of noise types, most notably GNSS.

In efforts to circumvent this limitation, this technical note elaborates an analytical method to generate an approximated PSD expressed as a set of power-laws defined in specific frequency intervals, starting from an Allan/Hadamard Variance (AVAR/HVAR) expressed a set of power-laws in the time domain.

The manuscript is sectioned as follows. \sref{sect_math} outlines the theoretical basis of the method for a set of cases: 1) the AVAR $\iff$ PSD for a single slope 2) the exact PSD $\Rightarrow$ AVAR when the former is a continuous function formed by a set of power laws valid in discrete  intervals 3) the novel proposed algorithm for ADEV/HDEV $\Rightarrow$ PSD when the former is composed by a set of power laws valid in discrete intervals. In \sref{sect_numerical} we show a set of numerical tests based on Allan variances of real clocks and we discuss the limits of our method: asymptotic behavior
(\sref{sect_infty}) and the passage to the continuous case (\sref{sect_cont}). Finally, in \sref{sect_concl} we draw the possible applications and conclusions of this work.

\section{Mathematical framework}\label{sect_math}
The mathematical model of timing signals and the applicability of the AVAR is briefly recalled hereby.\\
For a given clock, represented as an oscillator with nominal frequency $\nu_0$, we define the phase as
\be
\Phi(t) = 2 \pi \nu_0 t + \varphi(t) = 2 \pi \nu_0 \left[t+x(t)\right]
\ee
where $\varphi(t)$ is the random component and $x(t)$ is the time-error function  (i.e. the difference between the time of the clock and a reference "real" time $t$). The instantaneous  frequency is defined as
\be
\nu(t)=\frac{1}{2\pi} \frac{\der \Phi}{\der t}.
\ee
For specific applications, the fractional-frequency error 
\be
y(t) = \dot x(t) = \frac{\nu(t)}{\nu_0}-1 
\label{eqyps}
\ee
is more convenient.\\
{ From the above definitions, the relations among the (single-sided) phase, time, and frequency fluctuation power spectra ($S_\varphi(f),S_x(f)$ and $S_y(f)$, respectively) are 
\be
S_\varphi(f)=(2 \pi \nu_0)^2 S_x(f); \qquad S_y(f)=(2 \pi f)^2 S_x(f).
\ee
The autocorrelation function of the  fractional-frequency error is (for real data)
\be
R_y(t) = \lim_{T\to +\infty} \frac{1}{2 T} \int_{-T}^{+T} y(t') y(t'+t) \der t'
\ee
where $T\to +\infty$ indicates the passage from periodic to nonperiodic signals. For the Wiener-Khinchin theorem $R_y(t)$ and $S_y(f)$ are Fourier Transform/Inverse Fourier Transform couple
 \be
S_y(f) = \int_0^{+\infty} R_y(t) \cos (2 \pi f t) \der t; \qquad R_y(t) =  \int_0^{+\infty} S_y(f) \cos (2 \pi f t) \der f
\ee
(single-sided). Since measurements are spaced by an interval of time $\tau$ (i.e. integration time), we define the following estimators
\be
y_{n+1}(t,\tau) = \frac{y_n(t+\tau)-y_n(t)}{c_{n+1}/c_n}; \qquad n=1,...
\label{eqestim}
\ee
where
\be
y_1(t,\tau) = \frac{x(t+\tau)/\tau-x(t)/\tau}{ c_1}
\ee
and $c_n$ are numerical coefficients (see below).\\
The autocorrelation of $y_n(t,\tau)$ is
\be
R_{y,n}(t,\tau) = \lim_{T\to +\infty} \frac{1}{2 T} \int_{-T}^{+T} y_n(t',\tau) y_n(t'+t,\tau) \der t'; \qquad n=1,...
\label{eqryy1}
\ee
By inserting \eref{eqestim} into \eref{eqryy1} it can be demonstrated that (see e.g. \cite{vanvliet1982,greenhall1997})
\be
R_{y,n}(t,\tau) = \frac{2^{2n-2}}{c_n^2} \int_0^{+\infty} S_y(f) \cos (2 \pi f t) \frac{\sin^{2n} (\pi f \tau)}{(\pi f \tau)^2} \der f; \qquad n=1,...
\label{eqryy2}
\ee
and the coefficients $c_n$ are obtained by conventionally imposing that
\be
R_{y,n}(0,\tau)=R_{y,1}(0,\tau)\quad \forall\, n
\ee
in the case of white noise (i.e. $S_y(f)$=constant). This corresponds to \cite{makdissi2010}
\be
c_n^2 = c_1^2 \frac{2^{2n-2} \Gamma(n-1/2)}{\sqrt \pi\, \Gamma(n)} \qquad n=2,...
\label{eqcn}
\ee
assuming $c_1=1$ one finds $c_2^2=2$, $c_3^2=6$, $c_4^2=20$, $c_5^2=70$, etc.\\
By inserting \eref{eqcn} into \eref{eqryy2}   one obtains
\be
R_{y,n}(t,\tau) = \frac{\sqrt \pi \, \Gamma(n)}{\Gamma(n-1/2)} \int_0^{+\infty} S_y(f) \cos (2 \pi f t) \frac{\sin^{2 n}(\pi f \tau)}{(\pi f \tau)^2}\der f.
\label{1sidedhavar0}
\ee
Finally, we define $R_{y,2}(0,\tau)$
 as the Allan variance (AVAR) and $R_{y,3}(0,\tau)$
 as the Hadamard variance (HVAR) (\cite{cutler1965,vanvliet1982}). This latter, is able to deal with more divergent noise sources, characterized by power spectra $\propto f^\alpha$ with $-5 < \alpha \leq -2$ that cannot be handled by the AVAR \cite{riley2008}.\\
From \eref{eqestim}, the corresponding estimators are
\begin{align}
y_2(t,\tau)  &= \frac{y_1(t+\tau)- y_1(t)}{\sqrt {2}}=\frac{x(t+2 \tau)-2 x(t+\tau)+x(t)}{\tau \sqrt {2}}\\
y_3(t,\tau)  &=\frac{y_2(t+\tau)-y_2(t)}{\sqrt 3}=\frac{x(t+3 \tau)-3 x(t+2\tau)+3x(t+\tau)-x(t)}{\tau \sqrt{6}}.
\end{align}
Basing on \eref{1sidedhavar0} with $n=(2,3)$ the single-sided AVAR/HVAR $\sigma_y^2(\tau$) can be expressed as 
\be
\sigma_y^2(\tau) =  2 \int_{0}^\infty S_y (f)  q(\pi \tau f) \frac{\sin^4 (\pi \tau f) }{(\pi \tau f)^2} \, \der f;
\label{1sidedhavar}
\ee
where $q(\pi \tau f)=1$ for the AVAR and  $q(\pi \tau f)= 4/3 \sin^2 (\pi \tau f)$ for the HVAR.\\
The conditions for the convergence of the integral into \eref{1sidedhavar} are discussed in the following sections.

\subsection{Single slope case}
A power-law in the frequency domain, describing a power/amplitude spectral density (PSD/ASD), has a direct correspondence to a power-law in the integration time domain, describing an Allan variance/deviation (AVAR/ADEV), and vice versa {\em if and only if there is a single slope in the $]0,+\infty[$ $f$ and $\tau$ domains} \cite{burgoon1978} { (the lower limit is included if $\alpha\geq 0$)}. 
This univocal correspondence is easily demonstrated by substituting \eref{Sy} into \eref{1sidedhavar}.\\
By making the change $z=\pi \tau f$, \eref{1sidedhavar} becomes
\be
\sigma_y^2(\tau) =  2 h \frac{ I^\infty(\alpha)}{\pi^{\alpha+1}} \tau^{-\alpha -1}
\ee
where the integral
\be
I^\infty(\alpha) = \int_0^\infty q(z) \frac{\sin^4 z}{z^{2-\alpha}} \, \der z
\label{eqintI}
\ee
is is convergent if $-3 < \alpha < +1$ (AVAR case) and $-5<\alpha<+1$ (HVAR case).\\
Analogously, {if we define a  single-sloped AVAR/HVAR}
\be
\sigma^2_y(\tau)=B \tau^\mu
\ee
(where $B$ is constant), $h$ and $\alpha$ in \eref{Sy} are given by
\be
h=\frac{B}{2 \pi^\mu J^\infty(\mu) };\qquad \alpha=-\mu-1;
\label{eq:biun}
\ee
where we defined
\be
J^\infty(\mu)=I^\infty(-\mu-1)=\int_0^\infty q(z)\frac{\sin^4 z}{z^{\mu+3} }\, \der z.
\ee
The integral $J^\infty(\mu)$ is convergent if $-2<\mu<+2$ (AVAR case) and $-2<\mu<+4$ (HVAR case).\\
Therefore, there is a biunivocal correspondance between $(h,\alpha)$ coefficients of the PSD and the $(B,\mu)$ coefficients of the AVAR/HVAR given by \eref{eq:biun} QED.\\ 
Details about the functions $I^\infty(\alpha)=J^\infty(-\alpha-1)$ are reported in \aref{app_jfun}.\\
To deal with $\alpha \geq 1$ (or $\mu \leq-2$), an upper cutoff frequency $f_H$ to assure the convergence is introduced. In this case the integral contains oscillating terms which are usually neglected if we limit to $\tau \gg 1/(2\pi f_H)$. As a consequence of this approximation, different slopes in frequency domain can correspond to the same slope in time domain, making impossible to obtain the PSD from the AVAR. This is the case of the flicker and white phase modulation noises ($\alpha=$ +1 and +2, respectively) both of which correspond, neglecting oscillating terms, to an AVAR $\propto \tau^{-2}$. This problem can be avoided by introducing the modified Allan variance \cite{riley2008}.\\
In the following we will limit our analysis to the cases for which the integral of \eref{eqintI} is convergent.\\
\subsection{Several power laws case: exact formula to obtain AVAR from PSD}
Defining an arbitrary (i.e. not necessarily equally-spaced) set of $n$ coefficients $\alpha_i$ and $n-1$ frequencies $f_i$ with $f_1<f_2<...<f_{n-1}$, we may write the PSD as a continuous function $S_y(f)$ expressed as a set of power laws
\be
S_y(f)=   \begin{cases} h_1 f^{\alpha_1}  & \mbox{if $f< f_1$} \\ 
h_i f^{\alpha_i} & \mbox{if $f_{i-1}< f < f_i$, with $2\leq i \leq n-1$}\\ 
h_n f^{\alpha_n}  & \mbox{if $f> f_{n-1}$}.  \end{cases}
\label{eqsy}
\ee
In the log-log plane (here and below $\log=\log_{10}$ and $\ln=\log_e$)
\be
\log S_y(f) = \log h_i + \alpha_i \log f ; \qquad f_{i-1}< f < f_i.
\ee
Defining $Q_{asd}=\sqrt{S_y(f_1)}$ (i.e. the value of the ASD at $f=f_1$) we obtain $h_1=Q_{asd}^2 f_1^{-\alpha_1}$.\\
Since $S_y(f)$ is a continuous function we have
\be
h_{i+1}=h_{i} f_{i}^{\alpha_{i} - \alpha_{i+1}}; \qquad i=1,...,n-1.
\label{eqcontinuitypsd}
\ee
To pass from PSD to AVAR we insert \eref{eqsy} into \eref{1sidedhavar}
\be
\sigma_y^2(\tau) = 2 \sum_{i=1}^n \int_{f_{i-1}}^{f_i} h_i f^{\alpha_i} q(\pi \tau f) \frac{\sin^4 (\pi \tau f)}{(\pi \tau f)^2} \der f
\ee
where we defined $f_0=0$ and $f_n=\infty$. 
With the variable change $z=\pi \tau f$ we obtain
\be
\sigma_y^2(\tau) = 2 \sum_{i=1}^n \frac{h_i}{(\pi \tau)^{\alpha_i+1}}\int_{\pi \tau f_{i-1}}^{\pi \tau f_i} q(z) \frac{\sin^4 z }{z^{2-\alpha_i}} \, \der z.
\ee
Therefore, a PSD in the form of \eref{eqsy} exactly corresponds to an AVAR in the form
\be
\sigma^2_y(\tau) = \sum_{i=1}^n C_i(\tau) \tau ^ {-\alpha_i -1}
\label{eqavar0}
\ee
where 
\be
C_i(\tau) = \frac{2 h_i}{{\pi}^{\alpha_i+1}}\int_{\pi \tau f_{i-1}}^{\pi \tau f_i}q(z) \frac{\sin^4 z }{z^{2-\alpha_i}} \, \der z \qquad i=1,..., n
\label{eqavar1}
\ee
act as a weight function. In particular
\be
C_1(\tau) = \frac{2 h_1}{{\pi}^{\alpha_1+1}}\int_{0}^{\pi \tau f_1}q(z)\frac{\sin^4 z }{z^{2-\alpha_1}} \, \der z; \qquad C_n(\tau) = \frac{2 h_n}{{\pi}^{\alpha_n+1}}\int_{\pi \tau f_{n-1}}^{\infty}q(z) \frac{\sin^4 z }{z^{2-\alpha_n}} \, \der z.
\ee
For $\tau \rightarrow 0$ the $n$-th term of \eref{eqavar0} becomes the dominant one while the integral tends to $I^\infty(\alpha_n) $ while, for $\tau \rightarrow \infty$, the first term becomes dominant and the integral tends to $I^\infty(\alpha_1)$. Therefore,
\begin{align}
\label{eqtendtozero}
    \sigma_y^2(\tau\rightarrow 0) & \approx  \tau^{-\alpha_n-1} \frac{2 h_n}{\pi^{\alpha_n+1}}I^\infty(\alpha_n); \\
    \label{eqtendtoinf}
    \sigma_y^2(\tau \rightarrow \infty) & \approx \tau^{-\alpha_1-1}  \frac{2 h_1}{\pi^{\alpha_1+1}} I^\infty(\alpha_1). 
\end{align}
Independently of $n$, the asymptotic behaviors at $\tau\rightarrow 0$ and $\tau \rightarrow \infty$ are given by adopting the single-slope formula. 
Finally, we want to express the AVAR given by \eref{eqavar0} and \eref{eqavar1} in the form 
\be
\sigma_y^2(\tau) = \sum_{i=1}^n B_i(\tau) \tau^{\mu_i}
\ee
with $B_1 \tau^{\mu_1}$ at $\tau\approx 0$ and $B_n \tau^{\mu_n}$ for large $\tau$. Therefore, we need to reverse the indices by passing from $i$ to $n-i+1$
\be
\mu_i = -\alpha_{n-i+1}-1; \qquad i=1,..., n
\label{eqmui}
\ee
and, after substituting \eref{eqmui} into \eref{eqavar1}, we obtain
\be
B_i(\tau) = C_{n-i+1}(\tau) = \frac{2 h_{n-i+1}}{{\pi}^{\alpha_{n-i+1}+1}}\int_{\pi \tau f_{{n-i}}}^{\pi \tau f_{n-i+1}}q(z) \frac{\sin^4 z }{z^{2-\alpha_{n-i+1}}} \, \der z.
\label{eqgeneral}
\ee
It is easy to verify that for $\tau \to 0$ the AVAR reduces to $\approx B_1 \tau^{\mu_1}$, and for $\tau \to \infty$ the AVAR reduces to $\approx B_n \tau^{\mu_n}$.

\subsection{Several power laws case: approximated formula to obtain PSD from AVAR}\label{subsect_c}
We hereby propose the novel method to obtain a reasonable approximation of the PSD corresponding to a given AVAR/HVAR. 
Although this is not an exact conversion, one may always apply the exact formula \eref{eqgeneral} on the yielded PSD to validate the output against the original input AVAR/HVAR. \\
For a given set of $n$ values $\{\mu_1, ..., \mu_n \}$ and $n-1$ nodes $\{\tau_1,..., \tau_{n-1} \}$ with $\tau_1<\tau_2<...<\tau_{n-1}$ we define the AVAR as
\be
\sigma^2_y(\tau)=   \begin{cases} B_1 \tau^{\mu_1}  & \mbox{if $\tau< \tau_1$} \\ 
B_i \tau^{\mu_i} & \mbox{if $\tau_{i-1}< \tau < \tau_i$, with $2\leq i \leq n-1$}\\ 
B_n \tau^{\mu_n}  & \mbox{if $\tau> \tau_{n-1}$.}  \end{cases}
\label{eqavar}
\ee
As was for $S_y(f)$, in the log-log plane we have
\be
\log [\sigma_y^2(\tau)]_i = \log B_i + \mu_i \log \tau; \qquad \tau_{i-1}<\tau<\tau_i.
\ee
By defining $Q_{adev}=\sigma_y(\tau_1)$ we obtain $B_1=Q_{adev}^2 \tau_1^{-\mu_1}$ and the continuity constraint of $\sigma^2_y(\tau)$ leads to
\be
B_{i}\tau_i^{\mu_i}=B_{i+1} \tau_{i}^{\mu_{i+1}}; \qquad i=1,...,n-1.
\label{eqcontinuityb}
\ee
From \eref{eqtendtozero} and \eref{eqtendtoinf} we obtain $h_n$ and $h_1$, respectively, as functions of $B_1$ and $B_n$. These are the asymptotic behaviors at $f\rightarrow \infty$ and $f\rightarrow 0$, respectively.\\
The fundamental step is to extend this approach as to include the intermediate $h_i$. Therefore, the approximated PSD will resemble the form of \eref{eqsy}. From \eref{eqmui} we have
\be
\alpha_i = -\mu_{n-i+1}-1;\qquad i=1,...,n.
\label{eqalphai}
\ee
By defining
\be
 I_i^\infty= \int_{0}^{\infty}q(z) \frac{\sin^4 z }{z^{2-\alpha_i}} \, \der z; \qquad J_{i}^\infty  = \int_{0}^{\infty}q(z) \frac{\sin^4 z }{z^{3+\mu_{i}}} \, \der z \qquad  i=1,...,n;
\ee
we obtain
\be
I_i^\infty = J_{n-i+1}^\infty; \qquad J_i^\infty = I_{n-i+1}^\infty.
\ee
From \eref{eqgeneral}, and integrating between zero and $+\infty$ we obtain \footnote{\footnotesize The choice of integrating between zero and infinity has been done for the sake of simplicity. However, this imply that an exact convergence to the real PSD is impossible, also in the continuous case (see \sref{sect_cont}). The discrepancy being proportional to the degree of convexity/concavity of the AVAR.} 
\be
h_i  =  \frac{B_{n-i+1} }{2 J_{n-i+1}^\infty \pi^{\mu_{n-i+1}}}= \frac{B_{n-i+1} \pi^{\alpha_i+1}}{2 I_i^\infty}; \qquad i=1,...,n.
\label{eqhi}
\ee
Or, equivalently
\be
h_{n-i+1}  =  \frac{B_i }{2 J_i^\infty \pi^{\mu_i}}; \qquad i=1,...,n.
\label{eqhi2}
\ee
The final step is to obtain the {\em frequency nodes} $f_i$ ($1 \leq i \leq n-1$) where the $S_y(f)$ change-of-slopes occur. Using \eref{eqcontinuitypsd} (i.e. the continuity constraint for the PSD) 
we obtain
\be
f_i = \left(\frac{h_i}{h_{i+1}}\right)^{1/(\alpha_{i+1}-\alpha_i)}=\left[ \frac{B_{n-i+1} }  {B_{n-i}  } \frac{\pi^{\alpha_i+1}}{\pi^{\alpha_{i+1}+1}} \frac{I_{i+1}^\infty}{I_i^\infty}\right]^{1/(\alpha_{i+1}-\alpha_i)}; \qquad i=1,...,n-1.
\label{eqfi}
\ee
Using also \eref{eqcontinuityb} (i.e. the continuity constraint of the AVAR/HVAR), we obtain
\be
\frac{B_{n-i+1}}{B_{n-i}}= \tau_{n-i} ^{\mu_{n-i} - \mu_{n-i+1}} = \tau_{n-i}^{\alpha_{i}-\alpha_{i+1}} ; \qquad i=1,...,n-1.
\label{eqbbi}
\ee
Therefore by introducing \eref{eqbbi} into \eref{eqfi} we obtain the {\em frequency nodes}
\be
f_i =  \frac{1}{\pi \tau_{n-i} } \left[\frac{I_{i+1}^\infty}{I_i^\infty}\right]^{1/(\alpha_{i+1}-\alpha_i)} = \frac{1}{\pi \tau_{n-i} } \left[\frac{J_{n-i}^\infty}{J_{n-i+1}^\infty}\right]^{1/(\mu_{n-i+1}-\mu_{n-i})}; \qquad i=1,...,n-1
\label{eqfrnodes}
\ee
or
\be
    f_{n-i} = \frac{1}{\pi \tau_i } \left[\frac{J_i^\infty}{J_{i+1}^\infty}\right]^{1/(\mu_{i+1}-\mu_i)}; \quad i=1,...,n-1.
 \label{eqfrnodes2}
\ee

Hereinabove, the PSD is expressed as a set of power-law functions $h_i f^{\alpha_i}$ in the intervals $[f_{i-1},f_i]$ (i.e. as \eref{eqsy}) where $h_i$ coefficients are given by \eref{eqhi}, $f_i$ nodes by \eref{eqfrnodes} and $\alpha_i$ coefficients by \eref{eqalphai}. This represents a good approximation of the input AVAR/HVAR($B_i,\tau_i,\mu_i$) defined in \eref{eqavar}, as will be demonstrated in the following section.\\
It must be said that the problem of converting the intersection nodes $f \iff \tau$ was tackled by \cite{burgoon1978}, which was limited to the (single-node) 2-sloped case. Table 2 of \cite{burgoon1978}  reports a set of coefficients to convert the nodes for integer values of $\mu$ (including the cases $\mu=-3,-2$ that require the introduction of an upper cutoff  frequency).  
The other reported coefficients can be easily obtained by \eref{eqfrnodes} with $n=2$ and $\mu=-1,0,+1$. The conversion of the "nodes" is not trivial as it strongly depends on the values of the slopes. As pointed out by \cite{burgoon1978}, it can lead to large errors if, for example, one adopts $f_i=1/\tau_{n-1}$.\\
Furthermore, it should be noted that \cite{vernotte1993oscillator} describe the "multivariance method" to infer the noise coefficient set which best fits a set of time variance measurements in a weighted-least-squares sense. However, for high-fidelity results one must implement a large set of variances, and the frequency intervals where the coefficient sets are valid is not specified.

\section{Numerical tests}\label{sect_numerical}
Having defined the recipe for converting the AVAR to PSD, the efficacy of this conversion is hereby tested, considering two schemes:
\begin{enumerate}
\item AVAR$\to$PSD$\to$AVAR: given an arbitrary AVAR/HVAR in the form of \eref{eqavar}, by means of our algorithm in \eref{eqhi2} we generate the approximate PSD, and then compare the corresponding AVAR/HVAR with the original one.
\item PSD$\to$AVAR$\to$PSD: given an input PSD spectrum, by \eref{1sidedhavar} we calculate the corresponding AVAR and, using the method described in \sref{subsect_c}, we re-obtain the PSD and we compare it with the input one.
\end{enumerate}

\subsection{Real Allan deviations (Algorithm Fidelity in the domain?) }
In \fref{fig_clocks} we report some applications of the scheme AVAR$\to$PSD$\to$AVAR. We consider three AVARs relative to the following cases:
\begin{itemize}
    \item AccuBeat Ultra Stable Oscillator (USO)\footnote{\footnotesize \url{https://www.accubeat.com/uso}}, a high stability crystal quartz oscillator tailored for deep space exploration, and recently selected to fly onboard the ESA JUICE mission to the Galilean moons \cite{shapira2016};
    \item Orolia?s space-qualifed RAFS\footnote{\footnotesize Orolia datasheet: \url{https://safran-navigation-timing.com/product/rafs/}}, currently state-of-the-art rubidium clock selected by ESA to provide the frequency reference onboard the Galileo Second Generation navigation satellites (in this case we considered the HDEV).
\item A fictitious and willingly complex ADEV with $\mu=\{-3/2,$ $-1,$ $0,$ $+1,$ $+3/2\}$, $\tau=\{2^{-2},2,2^2,2^4\}$ and $Q_{adev}=10^{-14.5}$. This corresponds to $\alpha=\{-5/2,$ $-2,$ $-1,$ $0,$ $+1/2\}$, $f=\{4.473\times 10^{-6},$ $1.053\times 10^{-3},$ $0.180,$ $33.87\}$ and $Q_{asd}=10^{-11.71}$. 
\end{itemize}
In all cases we found a good agreement between the input and calculated Allan variances.

\begin{figure}[h!]
\begin{center}
\includegraphics[width=0.49\columnwidth]{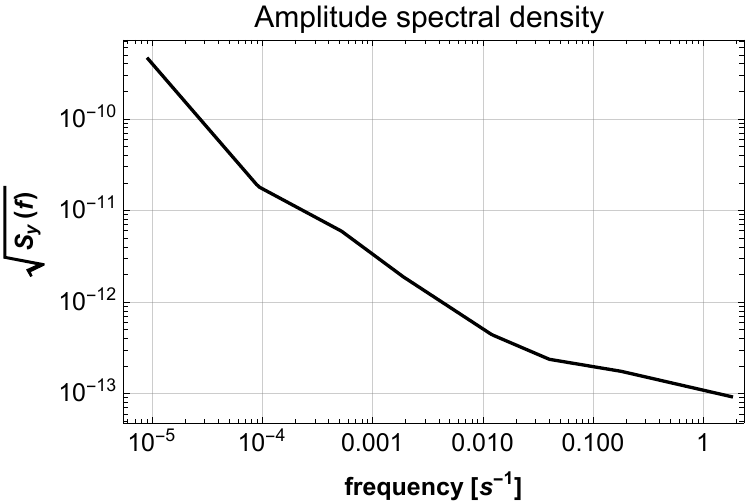}
\includegraphics[width=0.49\columnwidth]{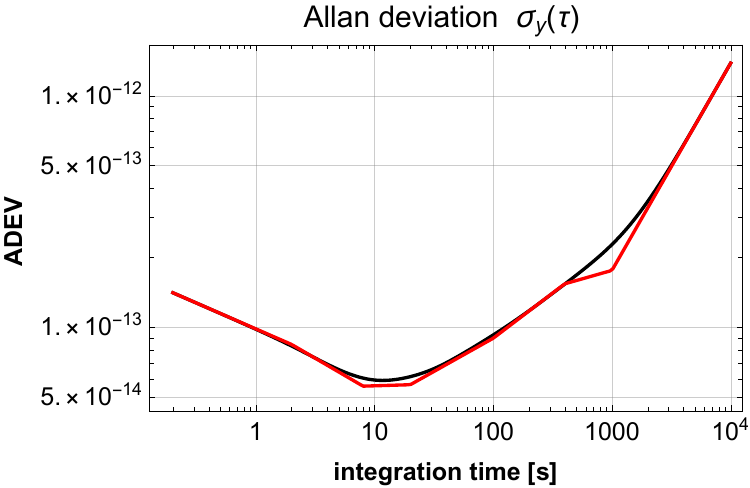}
\includegraphics[width=0.49\columnwidth]{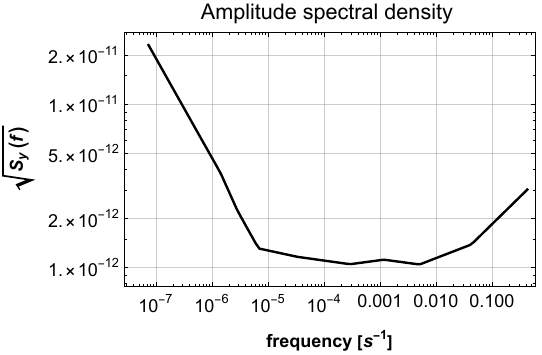}
\includegraphics[width=0.49\columnwidth]{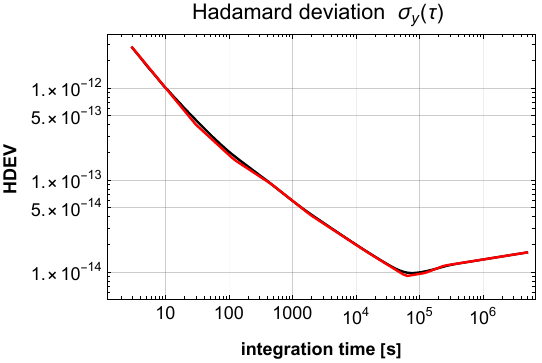}
\includegraphics[width=0.49\columnwidth]{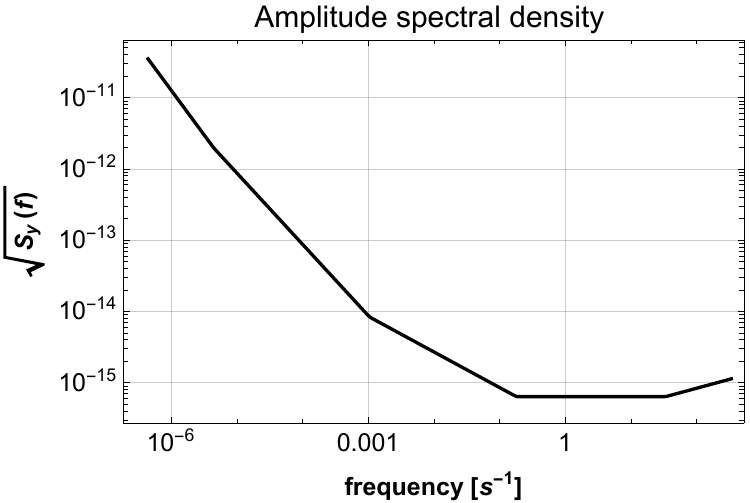}
\includegraphics[width=0.49\columnwidth]{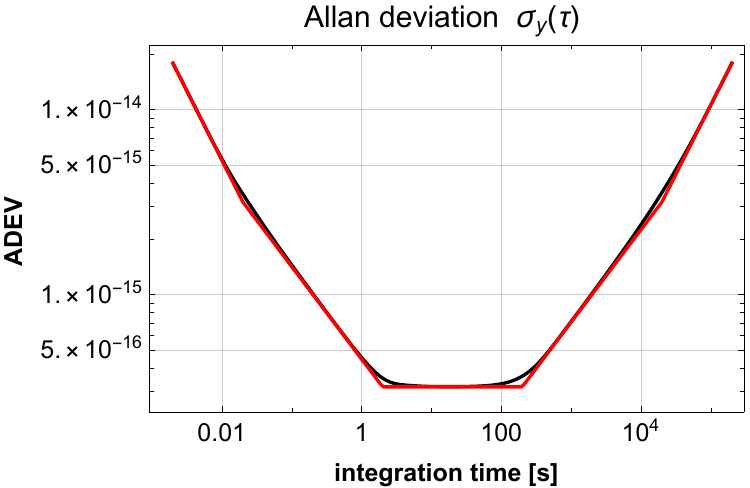}
\caption{\footnotesize Left: ASD obtained by the input ADEV (right, red lines). Right: input ADEV vs. ADEV (black line) corresponding to the ASD on the left. Top to bottom: AccuBeat, RAFS clocks (Hadamard deviation in this case). The last one been invented (see text for details).}
\label{fig_clocks}
\end{center}
\end{figure}
\clearpage
Moreover, we tested the scheme PSD$\to$AVAR$\to$PSD in the case of a Lorentzian frequency noise (i.e. $S_y(f)=(1+f^2)^{-1}$). Using \eref{1sidedhavar}, we numerically calculated the AVAR at 100 values of the integration time. We used this set of nodes and slopes to re-obtain the PSD using the method described above. In \fref{fig_Lor} (left panel) we compare the Lorentzian spectrum and the calculated one (red and black lines, respectively). The corresponding  ADEV is reported in the right panel.

\begin{figure}[h!]
\begin{center}
\includegraphics[width=0.49\columnwidth]{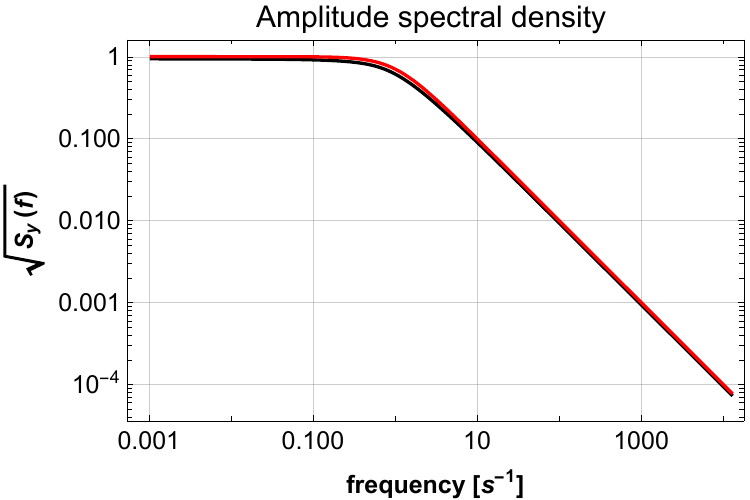}
\includegraphics[width=0.49\columnwidth]{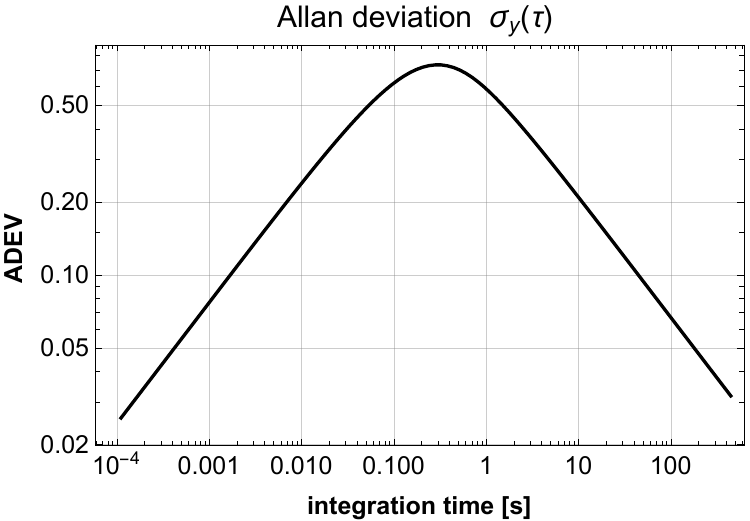}
\caption{\footnotesize Left: Lorentzian noise spectrum (red line) and the profile calculated basing on the ADEV reported on right panel (black line).}
\label{fig_Lor}
\end{center}
\end{figure}

{ 
Finally, we perform a more complete test. Its purpose is to show that the PSD obtained by the algorithm can be used to generate random noise whose ADEV agrees with the original one.\\ 
We consider three datasets, all relative to the trapped-ion Deep Space Atomic Clock (DSAC, that  completed its two-year mission in low-Earth orbit  on Sept. 18, 2021 \cite{burt2021}): 1) expected time fluctuations of DSAC-1 in flight (without other noise sources)  2) the same for DSAC-2 next generation clock and 3) ?raw? offsets, which includes measurement system noise, GPS noise, orbit determination noise, and DSAC-1 clock noise (this represents an upper bound for the clock itself).\\
For each dataset, starting from $N$ time-fluctuations data $x_i$ spaced by $\tau$, we directly obtain the AVAR as
\be
\sigma^2_{y}(\tau) 
= { 1 \over 2 (N-2) \tau^2 }
\sum_{n=1}^{N-2} ( {x}_{n+2} - 2x_{n+1} + x_{n} )^2
\label{avarnum}
\ee
for a set of integration times $\tau_j$.\\
By applying the algorithm described in \sref{subsect_c} we obtain the set of parameters  $(h_j,\alpha_j,f_j)$  relative to the PSDs $S_y(f)$ and $S_x(f)=S_y(f)/ (2 \pi f)^2$.\\
By generalising the noise-generating algorithms of \cite{timmer1995generating} to to an arbitrary PSD (as opposed to a single slope) we interpolate the $S_x(f)$ function with $n$ (even) frequencies between 0 and $f_n=1/\Delta t$, where $\Delta t$ is the desired spacing of the output data. We therefore generate the following complex vector $\ve F$ 
\be
X_j=  \frac{\sqrt{S_x(f_j)}}{2} \left[N_R^i(0,1)+ i N_j^i(0,1)\right]\quad \mbox{for }  j=1,n/2 +1; \qquad  X_{n/2+1+j} = X^*_{n/2-j+1} \quad \mbox{for } j=1,...,n/2
\ee
 where used the symmetry property of the discrete Fourier transform of real data. For this reason (Nyquist-Shannon theorem)  all information is contained below $f_{sup}=f_{n/2+1}=1/(2 \Delta t)$. Coefficients  $N_R^j(0,1),N_I^j(0,1)$ are two normally distributed random numbers and the factor 2 instead of $\sqrt 2$ at the denominator is necessary since  $S_x(f)$ is a single-sided spectrum.
It can be demonstrated that $<X_j^* X_j>=S_x(f_j)$.\\
Finally, we obtain the $n$-elements ($n$ is similar to the number of elements of the input datasets) vectors $\ve x$ and $\ve y$ of the simulated phase/time noises 
\be
\ve x = IDFT(\ve X) \sqrt{\frac{n-1}{\Delta t}}; \qquad  y_i = \frac{x_{i}-x_{i-1}}{\Delta t} \qquad i = 2,..., n 
\ee
where $IDFT$ is the  inverse discrete Fourier transform.\\
Finally, by applying \eref{avarnum} to the simulated $\ve y$ vector, we obtain the stability corresponding to the noise. In the left panel of \fref{fig_simdata}, the ADEVs from the input data (colored lines) are compared to the ADEVs of the noise. In all cases, a good agreement is found, indicating that the algorithm is able to deduce, from the input ADEVs, a faithful representation of their spectral contents (the deduced ASDs are reported in right panel). For an improved confidence at higher integration times, one may choose to compute the OADEV, albeit this entails a computational expense.

\begin{figure}[h!]
\begin{center}
\includegraphics[width=0.49\columnwidth]{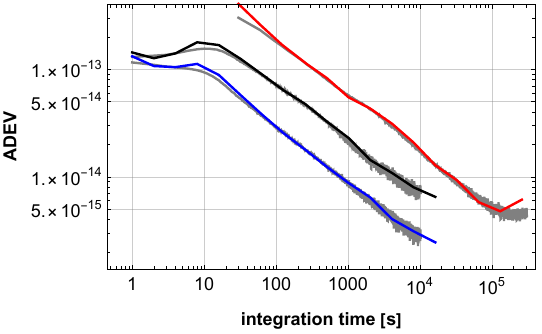}
\includegraphics[width=0.49\columnwidth]{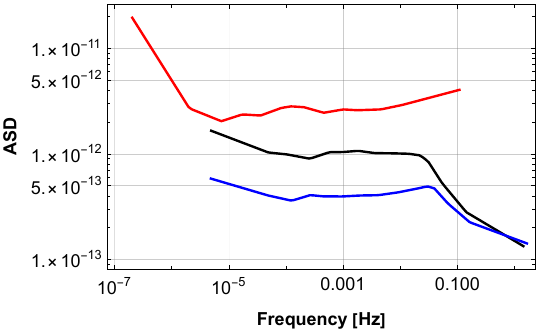}
\caption{\footnotesize Left: Allan deviations of the input data (black: DSAC-1, blue: DSAC-2, red: "raw" data) compared with the simulated time fluctuations noises (grey lines). Right:  ASDs estimated by the algorithm starting from the ADEVs of the input data.  Input data spacings are 1s (DSAC-1 and 2) and 30s ("raw" data).  }
\label{fig_simdata}
\end{center}
\end{figure}

}
\subsection{Asymptotic behaviour at zero and infinity}\label{sect_infty}
By inserting \eref{eqhi2} into \eref{eqgeneral} we obtain
\be
\hat \sigma_y^2(\tau) = \sum_{i=1}^n \frac{B_i \tau^{\mu_i}}{J_i^\infty}\int_{\pi \tau f_{n-i}}^{\pi \tau f_{n-i+1} } q(z) \frac{\sin^4 z}{z^{3+\mu_i}}  \, \der z; \qquad (f_0=0;\quad f_n=+\infty).
\label{eqsigmaapprox}
\ee
Therefore, the approximated AVAR/HVAR is in the form
\be
\hat \sigma_y^2(\tau) = \sum_{i=1}^n W_i(\tau) B_i \tau^{\mu_i} 
\ee
where
\be
W_i(\tau) = \frac{1}{J_i^\infty}\int_{\pi \tau f_{n-i}}^{\pi \tau f_{n-i+1} } q(z) \frac{\sin^4 z}{z^{3+\mu_i}}  \, \der z.
\ee
 By isolating the $i=1$ and $i=n$ terms into \eref{eqsigmaapprox} we obtain
\be
\hat \sigma_y^2(\tau) = \frac{B_1 \tau^{\mu_1}}{J_1^\infty}\int_{\pi \tau f_{n-1}}^{\infty } q(z) \frac{\sin^4 z}{z^{3+\mu_1}}  \, \der z+\sum_{i=2}^{n-1} \frac{B_i \tau^{\mu_i}}{J_i^\infty}\int_{\pi \tau f_{n-i}}^{\pi \tau f_{n-i+1} } q(z) \frac{\sin^4 z}{z^{3+\mu_i}} \, \der z+\frac{B_n \tau^{\mu_n}}{J_n^\infty}\int_{0}^{\pi \tau f_{1} } q(z) \frac{\sin^4 z}{z^{3+\mu_n}} \, \der z.
\ee
Independently of the number of frequency nodes (i.e. the size of $\Delta f_{n-i}$) when $\tau \rightarrow 0$, the calculated $\hat \sigma_y^2(\tau) $ cannot converge to the input AVAR/HVAR since $f_1$ and $f_{n-1}$ are not zero and infinity, respectively.\\
However, the integral into the first term tends to $J_1^\infty$, while others tend to zero in all cases. Therefore, $\hat \sigma^2(\tau)$ is asymptotic to $B_1 \tau^{\mu_1}$ for $\tau \to 0$. On the contrary, when $\tau \rightarrow +\infty$ the last integral tends to $J_n^\infty$ and the others tend to zero (because we are integrating in a part of the domain where the function is small). Therefore, $\hat \sigma^2(\tau)\rightarrow B_n \tau^{\mu_n}$.\\
In \fref{fig_clocks_prova1} we report the results for an input ADEV with uniformly randomly chosen $\mu$ parameters between -1.5 and 1.5  (red line). The asymptotic behavior at zero and infinity is always assured.\\
The {\em calculated } PSD (black line, left panel) corresponds to the ADEV in the right panel (black line). Note that the larger is the degree of concavity/convexity in the input ADEV, the larger are the discrepances.

\begin{figure}[h!]
\begin{center}
\includegraphics[width=0.49\columnwidth]{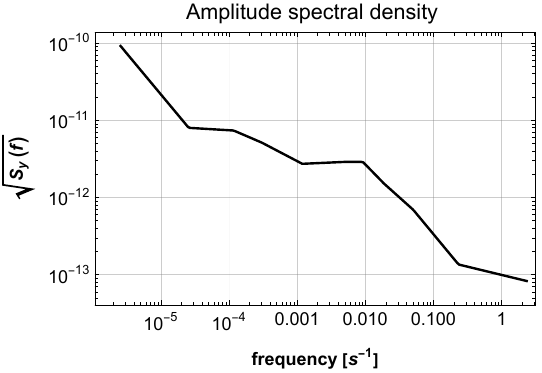}
\includegraphics[width=0.49\columnwidth]{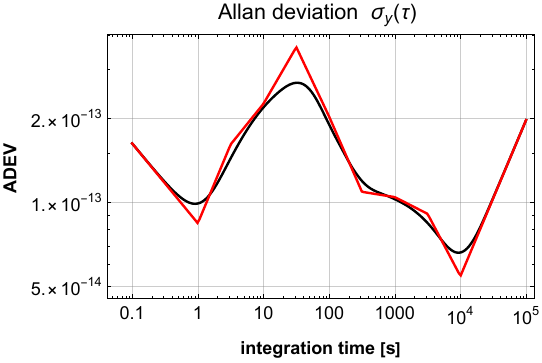}
\caption{\footnotesize Left: ASD obtained by the input ADEV (right, red lines). Right: input ADEV vs. reconstructed ADEV (black line) from ASD on the left. The ADEV profile has been invented (coefficients $\mu$ uniformly randomly chosen between -1.5 and +1.5).}
\label{fig_clocks_prova1}
\end{center}
\end{figure}

In \aref{app_2slopes} we demonstrate that if $n=2$ (i.e. two slopes only), the reconstructed AVAR/HVAR, in an interval around $\tau_1$, is above the input one if this latter is convex. It is below if the input function is concave.\\

\section{Continuous case}\label{sect_cont}
We will calculate the continuous version of the method described above (i.e.: when the number of sampling points tends to infinity but also the interval $[f_1,f_{n-1}]$ is expanded to  $[0,+\infty[$).\\
From \eref{eqavar0} and \eref{eqavar1}, the AVAR/HVAR corresponding to a PSD defined in terms of power-laws is 
\be
\hat \sigma^2_y(\tau) = \sum_{i=1}^n  \tau ^ {-\alpha_i -1} \frac{2 h_i}{{\pi}^{\alpha_i+1}}\int_{\pi \tau f_{i-1}}^{\pi \tau f_i}q(z) \frac{\sin^4 z }{z^{2-\alpha_i}} \, \der z \, \qquad i=1,..., n;
\label{eqfirst}
\ee
where $f_0=0$ and $f_n=+\infty$.
Since (for $x$ near to $a$)
\be
\int_a^x f(t) \, \der t \approx  f(a) (x-a) + \dots ,
\ee
by assuming an infinite number of frequency nodes, we get
\be
\sigma^2_y(\tau) = 2 \sum_{i=1}^{+\infty} h_i f_{i-1}^{\alpha_i} q(\pi \tau f_{i-1}) \frac{\sin^4 (\pi \tau f_{i-1}) }{( \pi \tau f_{i-1})^2} \,  \Delta f_{i-1}
\label{eqcont1}
\ee
where $\Delta f_{i-1}=f_i-f_{i-1}$. 
In the continuous case $f_i \to f$ 
(e.g.: a variable defined in the  $[0,+\infty[$ domain), $\alpha_i \to \alpha(f)$ and  $\Delta f_{i-1} \to \der f$. The summation becomes an integral and we re-obtain \eref{1sidedhavar} where $S_y(f) = h(f)$.\\
In the case of AVAR/HVAR, the passage to the continuous corresponds to $\tau_i \to \tau$, $\mu_i \to \mu(\tau)$, $J_i^\infty \to J^\infty[\mu(\tau)]$, $B_i\to B(\tau)$ and $\sigma_y^2(\tau) = B(\tau) \tau^{\mu(\tau)}$.\\  
By inserting \eref{eqhi} into \eref{eqfirst} and using \eref{eqalphai} we get
\be
\hat \sigma^2_y(\tau) = \sum_{i=1}^n  \tau ^ {\mu_{n-i+1}} \frac{B_{n-i+1} }{ J_{n-i+1}^\infty}  \int_{\pi \tau f_{i-1}}^{\pi \tau f_i}q(z) \frac{\sin^4 z }{z^{3+\mu_{n-i+1}}} \, \der z \, \qquad i=1,..., n.
\ee
The subscript $n-i+1$ indicates that $B, \mu$ and $J^\infty$ are relative to $\tau=\tau_{n-i+1}$ (e.g.: $B_{n-i+1}=B(\tau_{n-i+1})$). From \eref{eqfrnodes} we obtain
 \be
 \tau_{n-i+1} = \frac{1}{\pi f_{i-1} } \left[\frac{J_{i-1}^\infty}{J_i^\infty}\right]^{1/(\mu_i-\mu_{i-1})}; \qquad i=1,...,n-1.
\ee
Therefore, we can express $B, J^\infty$ and $\mu$ as functions of $f_{n-i}$.
Assuming infinite frequency nodes we get
\be
\hat \sigma^2_y(\tau) = \sum_{i=1}^{+\infty} \frac{B(f_{i-1}) }{ J^\infty(f_{i-1}) \pi^{\mu(f_{i-1})}} f_{i-1}^{-1-\mu(f_{i-1})} q(\pi \tau f_{i-1}) \frac{\sin^4 (\pi \tau f_{i-1}) }{( \pi \tau f_{i-1})^2} \,  \Delta f_{i-1}.
\label{eqcont2}
\ee
 When passing to the continuous case, we define
\be
Z(\mu_{i})=\lim_{\Delta \mu_{i} \to 0} \left[\frac{J_{i}^\infty}{J_{i+1}^\infty}\right]^{1/\Delta \mu_i}; \qquad \mbox{where } \Delta \mu_i = \mu_{i+1}-\mu_i.
\ee
The limit is an indeterminate form ($1^\infty$) and it is easy to demonstrate that
\be
Z(\mu) = \exp{\left[-\frac{\der \ln J^\infty(\mu)}{\der \mu}\right]}
\label{eq50}.
\ee
Note that the limit $\Delta \mu \to 0$ means that the input function must be not only continue but also "smooth" (i.e. also its first derivative must be continue).\\
Therefore, in the continuous case, \eref{eqfrnodes} and \eref{eqfrnodes2} become
\be
f(\tau')=\frac{Z[\mu(\tau')]}{\pi \tau'}.
\label{ftau}
\ee
If allowed (i.e. if and only if $f(\tau')$ is bijective), this relation can be inverted to calculate $\tau'(f)$.\\
Details about the function $Z(\mu)$ are reported in \aref{app_zfun}.\\
Passing to the continuous case, \eref{eqcont2} becomes
\be
\sigma^2_y(\tau) \approx  \int_0^{+\infty} \frac{B(f) }{ J^\infty[\mu(f)] \pi^{\mu(f)}} f^{-1-\mu(f)} q(\pi \tau f) \frac{\sin^4 (\pi \tau f ) }{( \pi \tau f)^2} \,  \der f  
\ee
the corresponding PSD is therefore 
\be
S_y(f) = \frac{B[\tau'(f)] }{2 J^\infty[\mu(\tau'(f))] \pi^{\mu[\tau'(f)]}} f^{-1-\mu[\tau'(f)]}
\label{eqfinsy}
\ee
where $\tau'(f)$ is obtained by inverting \eref{ftau}.\\
To invert \eref{ftau} and to apply \eref{eqfinsy}, we need to calculate $\mu(\tau)$ and $B(\tau)$ from a given input AVAR/HVAR $\sigma_y^2(\tau)$.\\
To this aim, we interpolate it with a set of power laws as in \eref{eqavar} with the same continuity constraint. We define a set of nodes $\{\tau_1,...,\tau_{n-1}\}$ and we pass to the continuous case
\be
\mu_{i}  = \frac{\log \sigma_y^2(\tau_i)-\log \sigma_y^2(\tau_{i-1})}{\log \tau_{i}-\log \tau_{i-1}} \quad \to \quad  \mu(\tau) = \frac{\der \log \sigma_y^2(\tau)}{\der \log \tau}\\
\label{eqmuf}
\ee
and
\be
\log B_{i} = \log \sigma_y^2(\tau_i)-\mu_{i} \log \tau_{i} \quad \to \quad \log B(\tau) = \log \sigma_y^2(\tau) -\mu(\tau) \log \tau.
\label{eqb0}
\ee
If the AVAR/HVAR is given as an analytical function, \eref{eqfinsy} provides the corresponding analytical PSD. It represents the limit of the (approximated) PSD described in \sref{subsect_c} when the input AVAR/HVAR is interpolated by a number of nodes that tends to infinity.\\
In \aref{app_contcheck} we report two checks of \eref{eqfinsy}.

\section{Conclusions}\label{sect_concl}

Depending on the scenario, most notably due to equipment availability or capability, it may be preferable to characterize clock and oscillator instabilities in time or frequency domains, hence the need to translate between the two. In this work we describe a simple algorithm to numerically compute an approximated power spectral density (PSD) corresponding to an Allan (or Hadamard) variance (AVAR/HVAR) given as an input. This may be applied directly if the Allan deviation is given in terms of an arbitrary set of joined power-laws defined in contiguous intervals of time, but we also report the formula to be used if the Allan deviation is expressed in terms of an analytical function. \\

The algorithm is summarized as follows:
\begin{enumerate}
    \item Express the input ADEV/HDEV in power-law form by selecting a set ($\tau_i, \mu_i)$ and computing $B_i$ by means of \eref{eqcontinuityb};
    \item Use \eref{eqhi} to compute the set $h_i$ from $B_i$;
    \item Translate the slopes to the frequency domain with \eref{eqalphai};
    \item Compute frequency nodes from \eref{eqfrnodes}.
\end{enumerate}

Having obtained the PSD, one may apply the exact formulation \eref{eqavar1} to reconstruct the input AVAR/HVAR. This validation exercise has demonstrated the applicability of the proposed algorithm for "spectralizing" a series of clocks affected by combinations of noises, is illustrated in \eref{fig_clocks}. Furthermore, its limitations are investigated by applying it to "extreme" (and irrealistic) Allan deviations.  The discrepancy between inverse-calculated ADEV/HDEV and the input ADEV/HDEV is in general proportional to the local degree of concavity/convexity, nonetheless, in all cases, the calculated PSD tends to converge to the real one at the extremes of the frequency domain (zero and infinity). 

Foreseeable uses of this algorithm are twofold. A coupling with well-established algorithms relying on PSD \cite{timmer1995generating, Kasdin1995DiscreteSO} would constitute a versatile and effective tool for generating multi-colored noise series in time domain, for addressing the deleterious effects that oscillator instabilities (standalone or embedded in a system with external disturbances) produce on the overall performance. In turn, this may help to optimize the operation of reference time scales for applications such as GNSS, where predictions of the time deviation for free-running clocks are required.

A second foreseeable use is the computation of the autocorrelation matrix for estimation filters which process non-gaussian observables, as in the context of spacecraft navigation. When using a single datatype for the estimation of orbit and clock parameters, as is prospected for the one-way navigation of the next-generation of deep-space probes \cite{ely2018using}, errors can inject in the clock estimates some residuals of a non-perfect orbit estimation. The "true" autocorrelation may therefore help to effectively disentangle orbit and clock parameters in the estimation filter, for an effective decoupling of transnational and time dynamics.


\appendix

\section{Coefficients of $I^\infty(\alpha)$ }\label{app_jfun}
The integrai $I^\infty(\alpha)$ can be analytically expressed as
\be
I^\infty_{ADEV}(\alpha) = \Gamma[\alpha-1] \sin\left(\frac{\alpha \pi}{2}\right) \times   2^{-1 - 2 \alpha} (1 - 2^{1 + \alpha})
 \label{eqj1}
\ee
and
\be
I^\infty_{HVAR}(\alpha) =  \Gamma[\alpha-1] \sin\left(\frac{\alpha \pi}{2}\right) \times 2^{-1 - 2 \alpha} \times 3^{
 1 - \alpha} ( 4\times 3^\alpha -2^\alpha  - 5 \times 6^\alpha)
 \label{eqj2}
\ee
where $\Gamma$ is the gamma function, which is undefined for non-positive integers. In the range of values of our interest ($-3<\alpha<1$, AVAR; $-5<\alpha<1$, HVAR), the functions reported into \eref{eqj1} and \eref{eqj2} are undefined at $\alpha=\{-2,-1,0\}$ and $\alpha=\{-4,-3,-2,-1,0\}$ (HDEV), respectively. However, it is easy to demonstrate that their domains can be extended to these points by analytical continuation.\\
Plots of $J^\infty(\mu)=I^\infty(-\alpha-1)$ functions for Allan (left) and Hadamard (right) deviations are reported in \fref{jfun}.\\
Coefficients of $I^\infty(\alpha)=J^\infty(-\mu-1)$ for some values of $\alpha$ (or $\mu=-\alpha-1$) for Allan and Hadamard deviations are reported in \tref{tabj3}. 

\begin{figure}[h!]
\begin{center}
\includegraphics[width=0.49\columnwidth]{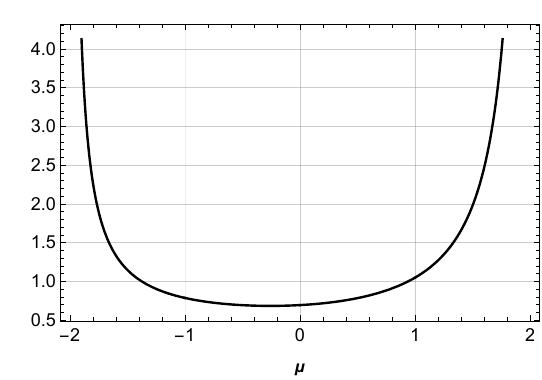}
\includegraphics[width=0.49\columnwidth]{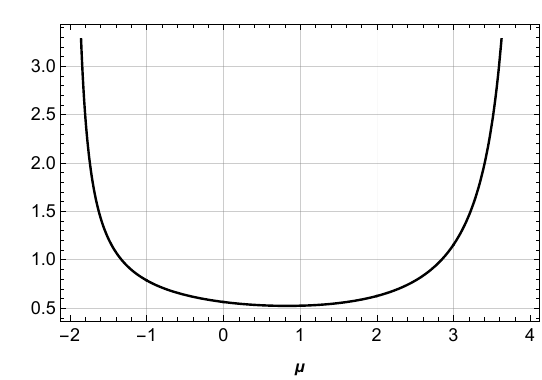}
\end{center}
\caption{\footnotesize Allan (left) and Hadamard (right) functions $J^\infty(\mu)$. The minimum is at: $\bar \mu=-0.25677$, $J(\bar \mu)=0.682881$ (Allan) and $\bar \mu=0.83161$, $J^\infty(\bar \mu)=3.13152$ (Hadamard).}
\label{jfun}
\end{figure}

\begin{table}
\be
\begin{array}{|l |l |l l |l l| }
\hline
\hline
\alpha & \mu & I^\infty_{AVAR}(\alpha) & & I^\infty_{HVAR}(\alpha) & \\[8pt]
\hline
 -5 & 4 & - & - & +\infty  & +\infty  \\
 -4 & 3 & - & - & \dfrac{11 \pi }{5} & 6.9115 \\
 -3 & 2 & +\infty  & +\infty & \log \left(\dfrac{1594323 \sqrt{3}}{65536}\right) & 3.74091 \\
 -2 & 1 & \dfrac{\pi}{3} & 1.0472 & \pi  & 3.14159 \\
 -1 & 0 & \log (2) & 0.693147 & \dfrac{3}{2} \log \left(\dfrac{256}{27}\right) & 3.37401 \\
 0 & -1 & \dfrac{\pi}{4} & 0.785398 & \dfrac{3 \pi }{2} & 4.71239 \\
 1 & -2 & +\infty & +\infty & +\infty  & +\infty  \\
 \hline
\end{array}
\ee
\caption{\footnotesize Coefficients of $I^\infty(\alpha)=J^\infty(-\mu-1)$ for integer values of $\alpha=-\mu-1$ for Allan and Hadamard deviations.}
\label{tabj3}
\end{table}

\section{Two slopes case ($n=2$)}\label{app_2slopes}
Here we demonstrate that, in the two-slopes case, $\hat \sigma_y^2(\tau_1)>B_1 \tau_1^{\mu_1}$ if and only if $\mu_2>\mu_1$ and vice versa.\\  
The reconstructed AVAR/HVAR is given by
\be
\hat \sigma_y^2(\tau) =  \frac{B_1 \tau^{\mu_1}}{J_1^\infty}\int_{\pi \tau f_1}^{\infty } q(z) \frac{\sin^4 z}{z^{3+\mu_1}} \, \der z+ \frac{B_2 \tau^{\mu_2}}{J_2^\infty}\int_0^{\pi \tau f_{1}} q(z) \frac{\sin^4 z}{z^{3+\mu_2}} \, \der z
\ee
where
\be
\pi \tau f_1=\frac{\tau}{\tau_1} \left(\frac{J_1^\infty}{J_2^\infty}\right)^{1/(\mu_2-\mu_1)}.
\ee
Since $B_2 =B_1 \tau_1^{\mu_1-\mu_2}$, we have
\be
\hat \sigma_y^2(\tau) =  B_1 \tau^{\mu_1} \left[\frac{1}{J_1^\infty}\int_{\pi \tau f_1}^{\infty } q(z) \frac{\sin^4 z}{z^{3+\mu_1}} \, \der z+ \frac{  (\tau/\tau_1)^{\mu_2-\mu_1}}{J_2^\infty}\left(1-\int_{\pi \tau f_{1}}^\infty q(z) \frac{\sin^4 z}{z^{3+\mu_2}} \, \der z\right)\right].
\ee
At $\tau=\tau_1$ 
\be
\hat \sigma_y^2(\tau_1) -B_1 \tau_1^{\mu_1}  = B_1 \tau_1^{\mu_1}   \int_0^{\pi \tau_1 f_1}  q(z) \frac{\sin^4 z}{z^{3}} \left(\frac{1}{J_2^\infty z^{\mu_2}}- \frac{1}{J_1^\infty z^{\mu_1}}\right)\, \der z. 
\ee
The integrand is equal to zero at 
\be
z = \left(\frac{J_1^\infty}{J_2^\infty}\right)^{1/(\mu_2-\mu_1)}=\pi \tau_1 f_1
\ee
(i.e.: the upper limit of integration). If $\mu_2>\mu_1$ the integrand is always positive, so the value of the reconstructed AVAR/HVAR is larger than the one of the input function (at the node $\tau_1$ but, for continuity, also in an interval that contains $\tau_1$). On the contrary, if $\mu_2<\mu_1$ the trial AVAR/HVAR is locally smaller than the input one. The discrepancy being proportional to the difference between $\mu_1$ and $\mu_2$ (in \fref{fig_2slopes} we report two examples with a big change in slope $\vert \mu_2-\mu_1\vert=3$). This is not, in general, true if $n>2$.

\begin{figure}[h!]
\begin{center}
\includegraphics[width=0.49\columnwidth]{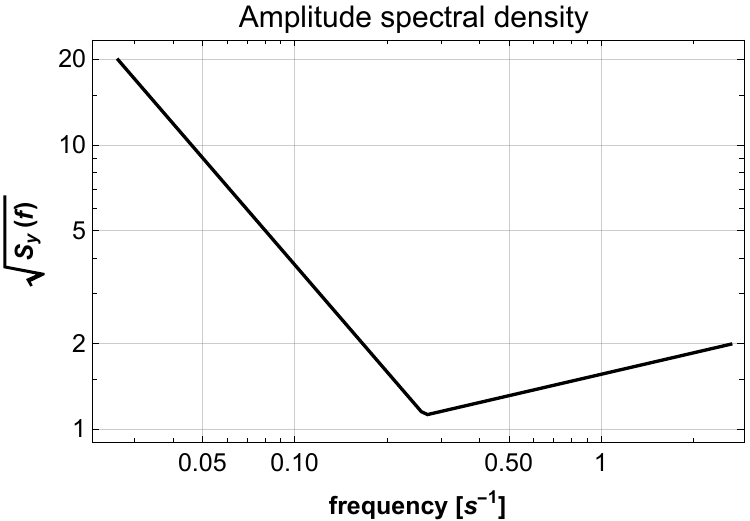}
\includegraphics[width=0.49\columnwidth]{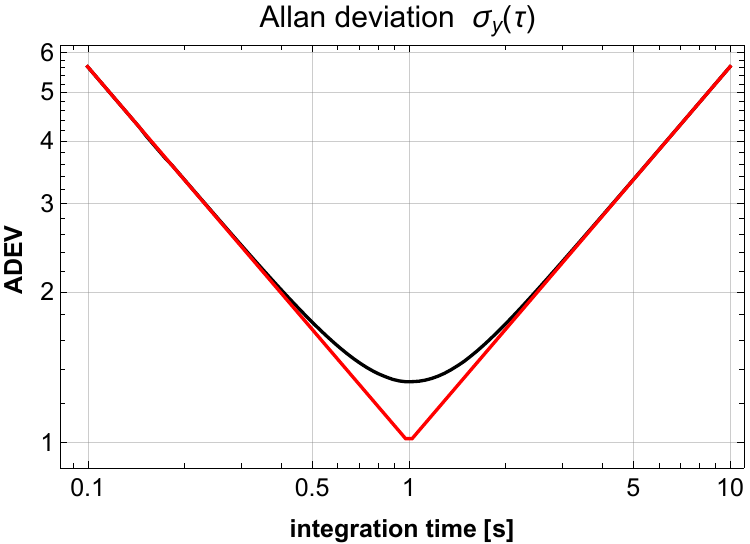}
\includegraphics[width=0.49\columnwidth]{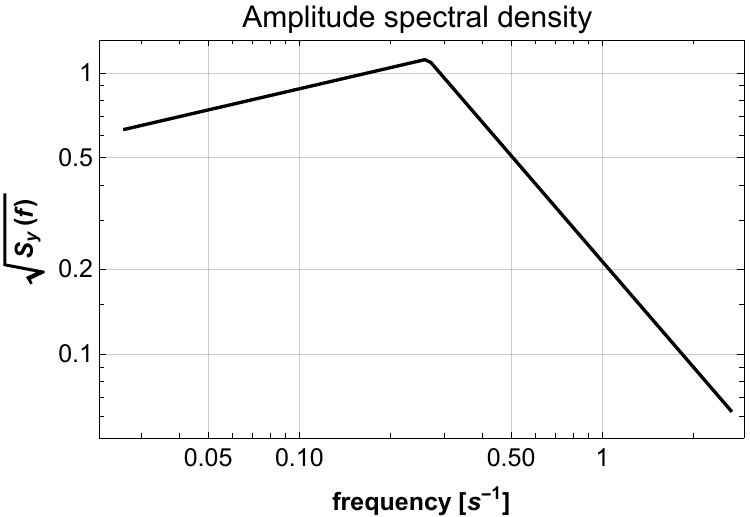}
\includegraphics[width=0.49\columnwidth]{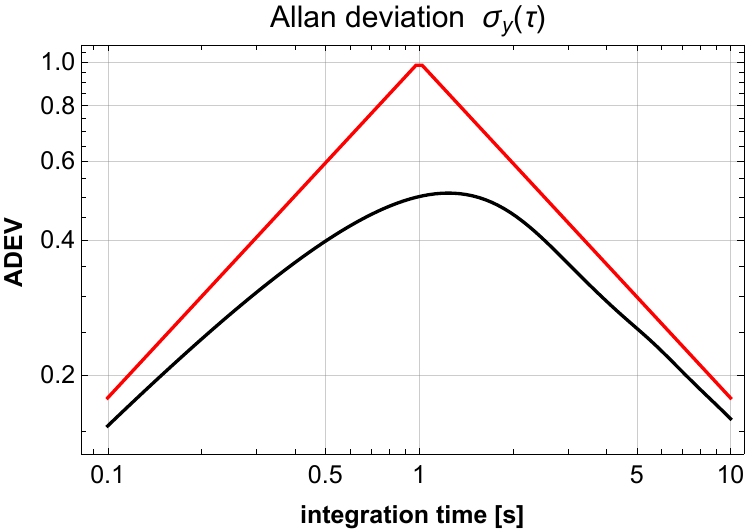}
\caption{\footnotesize Top ASD (left) and ADEV (right) for a 2-slopes case. The ADEV corresponding to the ASD on the left is above the input ADEV (red line) if $\mu_2>\mu_1$ (here $\mu_1=-1.5; \mu_2=+1.5$) and below otherwise (bottom panels, where $\mu_2=-1.5; \mu_1=+1.5$). }
\label{fig_2slopes}
\end{center}
\end{figure}

\clearpage

\section{The function $Z(\mu)$}\label{app_zfun}
From \eref{eq50}, the natural logarithms of $Z(\mu)$ are
\be
\ln [ Z^\infty_{AVAR}(\mu)] =\frac{\pi}{2}  \tan \left(\frac{\pi  \mu }{2}\right)+\psi (-\mu -2)+\left(\frac{1}{1-2^{\mu }}-2\right) \ln (2)
\ee
and
\be
\ln [ Z^\infty_{HVAR}(\mu)] = \frac{\pi}{2}  \tan \left(\frac{\pi  \mu }{2}\right)+\psi (-\mu -2)+\left(\frac{2^{\mu +3}}{-2^{\mu +3}+3^{\mu +1}+5}-1\right) \ln (2)+\frac{\ln (3)}{3^{-\mu -1}
   \left(2^{\mu +3}-5\right)-1}
\ee
where $\psi(x)=\Gamma'(x) / \Gamma(x)$ is the digamma function.\\
The function $Z(\mu)$ is plotted in \fref{zfun} for the Allan and Hadamard cases.\\
As was for $J^\infty(\mu)$, also $Z(\mu)$ must be extended to some points by analytical continuation.\\
For example, in the ADEV case the points are $\mu=\{-1,0,1\}$ where the limits are
\be
Z(-1)=e^{1-\gamma}; \qquad Z(0)=\frac{e^{3/2-\gamma }}{2 \sqrt{2}}; \qquad Z(1)=\frac{1}{8} e^{11/6-\gamma }
\ee
where $\gamma= 0.577215664...$ is the Euler?Mascheroni constant.\\

\begin{figure}[h!]
\begin{center}
\includegraphics[width=0.49\columnwidth]{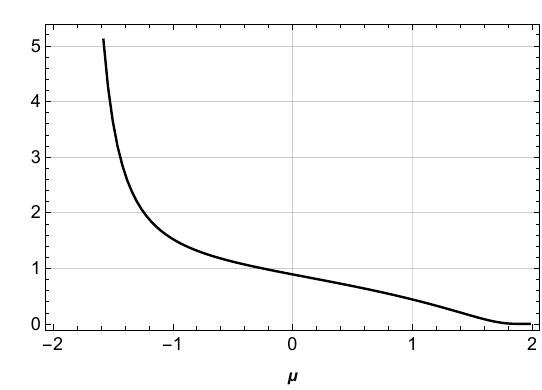}
\includegraphics[width=0.49\columnwidth]{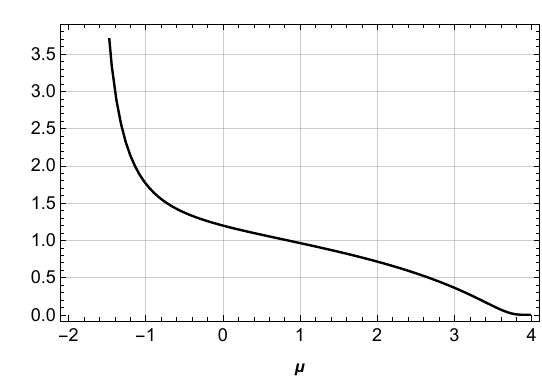}
\end{center}
\caption{\footnotesize Allan (left) and Hadamard (right) functions $Z(\mu)$.}
\label{zfun}
\end{figure}

\section{Check of the continuous case formula}\label{app_contcheck}
To check \eref{eqfinsy}, we assume an AVAR in the form
\be
\sigma_y^2(\tau)=\sum_{i=1}^N b_i \tau^{\mu_i}
\ee
and the corresponding PSD can be  analytically calculated as
\be
S_y(f) = \sum_{i=1}^N\frac{b_i}{2 \pi^{\mu_i} J^\infty(\mu_i)} f^{-\mu_i-1}.
\label{eq_theopsd}
\ee 
We compare the PSD generated by \eref{eqfinsy} with \eref{eq_theopsd}. We consider a "realistic" case: $N=3$, $b_i=1 \,\forall i$ and $\mu_i=\{-1,0,1\}$. The corresponding PSD is
\be
S_y(f)=2+\frac{1}{2 f \ln 2}+\frac{3}{2 \pi ^2 f^2}.
\label{eq_psdtheo3}
\ee
By \eref{eqmuf} and \eref{eqb0} we obtain
\be
\mu(\tau) = \frac{\tau^2-1}{\tau^2+\tau+1}; \qquad B(\tau) = \tau^{\frac{\tau+2}{\tau^2+\tau+1}-2} \left(\tau^2+\tau+1\right).
\label{eqmufb0calc}
\ee
In \fref{fig_clocks_prova2} (left panel) we compare the analytical ASD ($\sqrt{S_y(f)}$ from \eref{eq_psdtheo3}, black line) with the one calculated by using \eref{eqfinsy} where $\mu(\tau)$ and $B(\tau)$ are given by \eref{eqmufb0calc} (red line). In the right panel we report the input ADEV ($\sigma_y(\tau)=\sqrt{\tau+1+1/\tau}$).\\
Finally, we consider the case  $N=2, b_i=1\, \forall i$ and $\mu_1=-1.9$, $\mu_2=+1.9$. In \fref{fig_clocks_prova3} we report the result: the discrepancy is due to the strong difference between $\mu_1$ and $\mu_2$ (since $\mu$ must be between -2 and 2, we are considering an extremely large degree of convexity).
\begin{figure}[h!]
\begin{center}
\includegraphics[width=0.49\columnwidth]{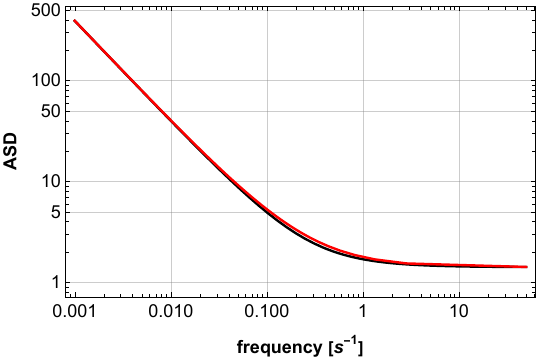}
\includegraphics[width=0.49\columnwidth]{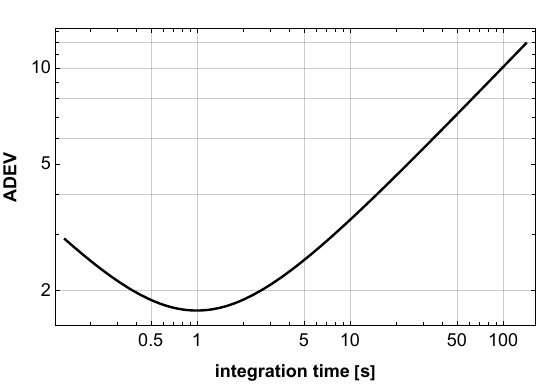}
\caption{\footnotesize Left: comparison between the analytical ASD (black line, i.e.: the square root of \eref{eq_psdtheo3}) and the calculated one (red line). Right: the input ADEV (i.e.: $\sigma_y(\tau)=\sqrt{\tau+1+1/\tau}$).}
\label{fig_clocks_prova2}
\end{center}
\end{figure}
\begin{figure}[h!]
\begin{center}
\includegraphics[width=0.49\columnwidth]{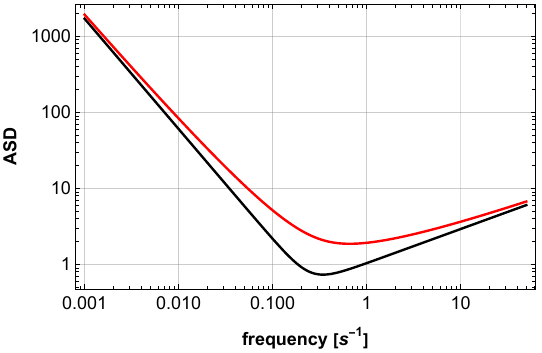}
\includegraphics[width=0.49\columnwidth]{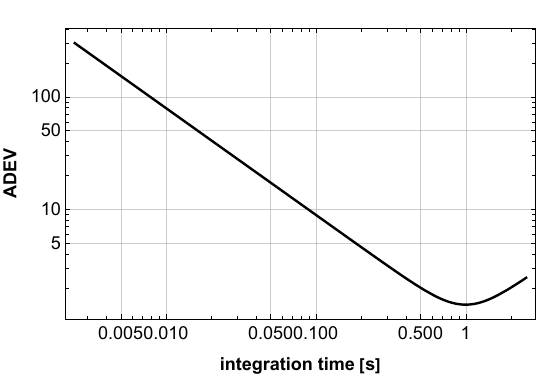}
\caption{\footnotesize The same as \fref{fig_clocks_prova2} but in the case $N=2$ and $\mu_1=-1.9$ and $\mu_2=+1.9$ (see text for details).}
\label{fig_clocks_prova3}
\end{center}
\end{figure}
\clearpage

%


\end{document}